\documentclass[sigconf]{acmart}

\AtBeginDocument{%
  }

\usepackage{algorithmic}
\usepackage{algorithm}

\setcopyright{acmcopyright}
\copyrightyear{2022}
\acmYear{2022}
\acmDOI{XXXXXXX.XXXXXXX}

\acmConference[UbiComp/ISWC '22 Adjunct]{UbiComp-ISWC '22: Adjunct Proceedings of the 2022 ACM International Joint Conference on Pervasive and Ubiquitous Computing and Proceedings of the 2022 ACM International Symposium on Wearable Computers}{September 11--15, 2022}{Atlanta, USA}
\acmBooktitle{Adjunct Proceedings of the 2022 ACM International Joint Conference on Pervasive and Ubiquitous Computing and Proceedings of the 2022 ACM International Symposium on Wearable Computers (UbiComp/ISWC '22 Adjunct), September 11--15, 2022, Atlanta, USA}
\acmPrice{15.00}
\acmISBN{978-1-4503-XXXX-X/XX/XX}

\begin{document}

\title{Privacy-Preserving Cooperative Visible Light Positioning for Nonstationary Environment: A Federated Learning Perspective}
\renewcommand{\shorttitle}{Privacy-Preserving Cooperative VLP for Nonstationary Environment: A Federated Learning Perspective}

\author{Tiankuo Wei}
\email{weitk@stu.xmu.edu.cn}
\affiliation{%
	\institution{Xiamen University}
	\city{Xiamen}
	\state{Fujian}
	\country{China}}

\author{Sicong Liu}
\authornote{Corresponding author.}
\email{liusc@xmu.edu.cn}
\affiliation{%
	\institution{Xiamen University}
	\city{Xiamen}
	\state{Fujian}
	\country{China}}

\renewcommand{\shortauthors}{Wei et al.}

\begin{abstract}
  Visible light positioning (VLP) has drawn plenty of attention as a promising indoor positioning technique. However, in nonstationary environments, the performance of VLP is limited because of the highly time-varying channels. To improve the positioning accuracy and generalization capability in nonstationary environments, a cooperative VLP scheme based on federated learning (FL) is proposed in this paper. Exploiting the FL framework, a global model adaptive to environmental changes can be jointly trained by users without sharing private data of users. Moreover, a Cooperative Visible-light Positioning Network (CVPosNet) is proposed to accelerate the convergence rate and improve the positioning accuracy. Simulation results show that the proposed scheme outperforms the benchmark schemes, especially in nonstationary environments.
\end{abstract}


\maketitle

\section{Introduction}
Nowadays, demands for indoor positioning systems are dramatically increasing due to the ongoing explosive growth of location based services. Unlike the outdoor environment that has been covered by the Global Navigation Satellite System (GNSS), indoor positioning is still a challenge due to the complicated environment \cite{Zafari2019COMSTPOS}. Recently, with the advent of visible light communication (VLC) and the ubiquitous deployment of light-emitting diodes (LEDs), visible light positioning (VLP) has drawn plenty of attention from both academia and industry \cite{Matheus2019VLC,Yang2016VLC,Yang2017TOBVLC}. It has been envisioned as a promising indoor positioning technique, because of the low hardware cost, high accuracy, unregulated spectrum, and low electromagnetic interference \cite{Luo2017COMSTVLP}.

Most of the research on VLP investigates the characteristics of received optical signals, such as received signal strength (RSS), time of arrival (TOA), and angle of arrival (AOA) \cite{Hass2018COMSTVLP,Sicong2015IEICE}. These characteristics are measured and mapped to the location coordinates with the positioning algorithms, such as triangulation and fingerprinting. Over the past few years, learning-based methods have been introduced into indoor positioning to further improve the positioning accuracy. For example, Van \emph{et al.} \cite{Van2017IETKNN} proposed an RSS-based method using weighted $k$-Nearest Neighbors ($k$-NN), which can achieve better accuracy than trilateration. Arfaoui \emph{et al.} \cite{Hass2021JSACVLP} proposed an RSS-based fingerprinting approach utilizing multiple layer perceptron (MLP) and convolutional neural network (CNN) to estimate both the position and orientation of the user equipment (UE). Lin \emph{et al.} \cite{Lin2020CLVLP} extracted the location information from the characteristic of channel impulse response with four structures of deep neural networks.

However, the performance of the learning-base methods may degrade in nonstationary environments, because of the time-varying channel property caused by obstacle movement, ambient light interference, device aging, etc. To improve the positioning accuracy and generalization capability in nonstationary environments, cooperative localization methods have been investigated. For example, Keskin \emph{et al.} \cite{Keskin2018VLP} designed a novel architecture that facilitates communications among VLC receivers for cooperative localization, and proposed a location estimation algorithm suitable for decentralized implementation. Yang \emph{et al.} \cite{Yang2019OEVLP} proposed a cooperative VLP system with the sum-product algorithm to achieve accurate positioning in the obstacle blocking situation. However, the position-related knowledge of the users remains to be fully extracted to further improve the positioning performance. Moreover, privacy data of users should be well protected, as the location data usually contains much private information.

Therefore, in this paper, we introduce the federated learning (FL) \cite{Jakub2016FLArxiv,YangQiang2019FL} framework and propose a privacy-preserving cooperative VLP scheme for nonstationary environments. Specifically, we design a cooperative VLP framework, where UEs participate in data collection and model training. Based on this framework, we propose an FL-based VLP algorithm aiming at jointly constructing an adaptive position estimation model without revealing their local data to the centralized server. To further improve the performance, we propose a Cooperative Visible-light Positioning Network (CVPosNet) to accelerate the convergence rate and improve the positioning accuracy.

The remainder of this paper is organized as follows. The system model of the indoor VLP is presented and the positioning problem is formulated in Section \ref{sec:systemModel}. The proposed cooperative VLP scheme is introduced in Section \ref{sec:proposedMethod}. The simulation results are reported and discussed in Section \ref{sec:simulationResults}, followed by the conclusions in Section \ref{sec:conclusion}.

\section{System Model and Problem Statement} \label{sec:systemModel}

\subsection{System Model of Indoor Visible Light Positioning}
\begin{figure}[!t]
	\centering
	\includegraphics[width=0.4\textwidth]{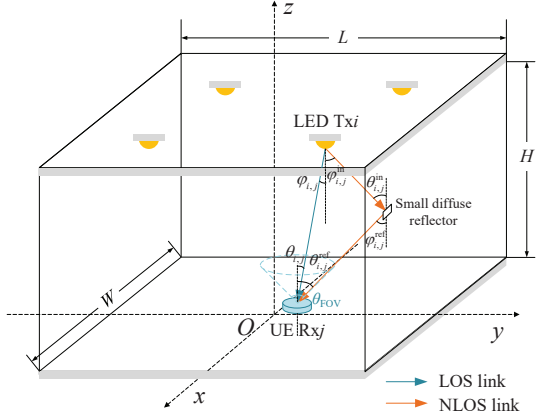}
	\caption{The channel model of the indoor VLP system.}
	\label{fig:systemModel}
\end{figure}

We consider an indoor VLP system as illustrated in Figure \ref{fig:systemModel}, which is deployed in a room with the size of $L \times W \times H$, where $N_\mathrm{t}$ LEDs are attached at the ceiling as positioning anchors, with the position coordinates of $\mathbf{a}_i=[x_i^\mathrm{a},y_i^\mathrm{a},z_i^\mathrm{a}]$. Without loss of generality, the optical wireless channel of both the light-of-sight (LOS) link and the non-light-of-sight link with once reflection on the walls are considered \cite{Komine2004VLCTCE}, and the optical channel gain between the $i$-th LED and the $j$-th UE located at $\mathbf{c}_j=[x_j,y_j,z_j]$ is given by
\begin{equation}
	H_{i, j}=H_{i, j}^{\mathrm{LOS}}+H_{i, j}^{\mathrm{NLOS}},
\end{equation}
where $ H_{i, j}^{\mathrm{LOS}} $ and $ H_{i, j}^{\mathrm{NLOS}} $ represents the channel direct current (DC) gains of the LOS and NLOS links, respectively.

Assuming that the LEDs follow a Lambertian emission pattern, the channel gain of the LOS link can be expressed as
\begin{equation}
	H_{i, j}^{\mathrm{LOS}}=\left\{\begin{array}{cc}
		\frac{(m+1) A_\mathrm{PD} \cos ^{m}\left(\varphi_{i, j}\right) \cos \left(\theta_{i, j}\right) T_\mathrm{s} g\left(\theta_{i, j}\right)}{2 \pi d_{i, j}^{2}}, & 0 \leq \theta_{i, j} \leq \theta_\mathrm{FOV}, \\
		0, & \theta_{i, j}>\theta_\mathrm{FOV},
	\end{array}\right.
\end{equation}
where $d_{i, j}=\left\|\mathbf{c}_{j}-\mathbf{a}_{i}\right\|_{2}$ is the Euclid distance between the $i$-th LED and the $j$-th UE. $m$ is the Lambertian order, $A_\mathrm{PD}$ is the effective physical area of the PD. $T_\mathrm{s}$ and $g\left(\theta_{i, j}\right)$ are the gain of the optical filter and the optical concentrator of the PD, respectively. $\theta_\mathrm{FOV}$ is the field-of-view (FOV) angle of the PD. $\varphi_{i, j}$ and $\theta_{i, j}$ are the angles of irradiance and incidence with respect to the normal direction, respectively. In addition, the Lambertian order is given by $m=-\ln 2 / \ln (\cos (\varphi_{1 / 2}))$, where $\varphi_{1/2}$ is the half-power angle of the LED. And the gain of the optical concentrator $g\left(\theta_{i, j}\right)$ can be expressed as
\begin{equation}
	g\left(\theta_{i, j}\right)=\left\{\begin{array}{cc}
		\frac{n^{2}}{\sin ^{2}(\theta_\mathrm{FOV})},  & 0 \leq \theta_{i, j} \leq \theta_\mathrm{FOV}, \\
		0, & \theta_{i, j}>\theta_\mathrm{FOV},
	\end{array}\right.
\end{equation}
where $n$ denote the refractive index of the optical concentrator.

The NLOS link is modeled as a diffusion channel, where the wall is segmented into many small diffuse reflectors with the area of $A_\mathrm{ref}$ and the average diffuse reflectance of $\bar{\rho}$. The NLOS channel gain can be approximately calculated as
\begin{equation}
	H_{i, j}^{\mathrm{NLOS}}=\left\{\begin{array}{cc}
		\sum_{\mathrm{Wall}}  \bar{\rho} h_{i, j}^{\mathrm{in}} h_{i, j}^\mathrm{ref}, & 0 \leq \theta_{i, j} \leq \theta_\mathrm{FOV}, \\
		0, & \theta_{i, j}>\theta_\mathrm{FOV}.
	\end{array}\right.
\end{equation}
Specifically, $h_{i, j}^{\mathrm{in}}$ and $h_{i, j}^{\mathrm{ref}}$ are the channel gains of incidence and reflection light given by
\begin{equation}
		h_{i, j}^\mathrm{in} = \frac{A_\mathrm{ref}(m+1)}{2 \pi (d_{i, j}^{\mathrm{in}})^2} \cos ^{m}\left(\varphi_{i, j}^\mathrm{in}\right) \cos \left(\theta_{i, j}^\mathrm{in}\right),
\end{equation}
\begin{equation}		
		h_{i, j}^\mathrm{ref} = \frac{A_\mathrm{PD}}{2 \pi (d_{i, j}^{\mathrm{ref}})^2} \cos \left(\varphi_{i, j}^\mathrm{ref}\right) \cos \left(\theta_{i, j}^\mathrm{ref}\right),
\end{equation}
where $d_{i, j}^{\mathrm{in}}$ represents the distance between the $i$-th LED and the diffuse reflector, and $d_{i, j}^{\mathrm{ref}}$ represents the distance between the diffuse reflector and the $j$-th PD. $\varphi_{i, j}^\mathrm{in}$ and $\theta_{i, j}^\mathrm{in}$ are the angle of irradiance and incidence of the incident light, respectively. $\varphi_{i, j}^\mathrm{ref}$ and $\theta_{i, j}^\mathrm{ref}$ are the angle of irradiance and incidence of the reflective light, respectively.

Thus, the received electronic power from the $j$-th LED is given by
\begin{equation}
	P_{i, j} = \left[R_\mathrm{p} P_{i}^\mathrm{t}\left(H_{i, j}^\mathrm{LOS} + H_{i, j}^\mathrm{NLOS}\right)\right]^{2} + \sigma_\mathrm{n}^{2},
\end{equation}
where $P_{i}^\mathrm{t}$ is the optical power emitted by the $i$-th LED, $R_\mathrm{p}$ is the responsivity of the PD. $\sigma_\mathrm{n}^{2} = \sigma_\mathrm{shot}^{2} + \sigma_\mathrm{thermal}^{2}$ is the noise variance composed of two components of the shot noise and the thermal noise, which are given by
\begin{equation}
	\sigma_{\mathrm{shot}}^{2}=2 q R_\mathrm{p} P_{i, j} B+2 q I_\mathrm{bg} I_{2} B,
\end{equation}
\begin{equation}
	\sigma_{\mathrm{thermal}}^{2}=\frac{8 \pi k T_\mathrm{K}}{G} \eta A I_{2} B^{2}+\frac{16 \pi^{2} k T_\mathrm{K} \Gamma}{g_\mathrm{m}} \eta^{2} A^{2} I_{3} B^{3},
\end{equation}		
where $q$ is the electronic charge, and $k$ is the Boltzmann constant. $I_\mathrm{bg}$ is the background current of the ambient light. $B$ is the noise bandwidth.  $T_\mathrm{K}$ is the absolute temperature. $G$ is the open-loop voltage gain. $\eta$ is the fixed capacitance of the PD. $\Gamma$ and $g_\mathrm{m}$ are the channel noise factor and the transconductance of the field effect transistors, respectively. $I_2$ and $I_3$ denote the noise bandwidth factors, which are defined as 0.562 and 0.0868, respectively \cite{Tang1996VLCNoise}. Finally, the received power vector of the $j$-th UE can be modeled as $\mathbf{p}_j=[P_{1,j},P_{2,j},\cdots,P_{N_\mathrm{t},j}]$.

\subsection{Problem Statement}

As mentioned above, the received power vector $\mathbf{p}_j$ is dependent on the channel gain $H_{i,j}$, which is controlled by the position-related parameters. Thus, the correlation between $\mathbf{c}_j$ and $\mathbf{p}_j$ can be exploited to estimate the location coordinate of the UE. Then, the positioning task can be mathematically formulated as
\begin{equation}
	\min _{\mathbf{w}, \mathcal{M}(\cdot)} \mathbb{E}_{\mathbf{c}_{j} \in \mathbb{C}} \left\| \hat{\mathbf{c}}_{j} - \mathbf{c}_{j} \right\|_{2}^{2},
\end{equation}
where $\hat{\mathbf{c}}_{j}=\mathcal{M}(\mathbf{p}_{j},\mathbf{w})$ denotes the estimated position coordinate obtained by a model $\mathcal{M}(\cdot)$ with the learnable parameters of $\mathbf{w}$. $\mathbb{C}$ is the set of the positions in the whole indoor environment lighten by the LEDs. To solve this problem, on the one hand, an efficient model $\mathcal{M}(\cdot)$ is needed to extract the position-related features and map them into the label space of location coordinates. On the other hand, it is also important to find a set of optimal parameters $\mathbf{w}^*$ adaptive to the indoor environment, which is more challenging in nonstationary environments, because of the highly time-varying channel condition caused by the environmental changes, such as ambient light interference, obstruction movement, and aging hardware. To improve the positioning accuracy and robustness in nonstationary environments, in this paper, an FL-based cooperative VLP scheme is proposed, which is introduced in detail in Section \ref{sec:proposedMethod}.

\section{Proposed Cooperative VLP Scheme} \label{sec:proposedMethod}

\subsection{Cooperative VLP Framework}

\begin{figure}[!t]
	\centering
	\includegraphics[width=0.45\textwidth]{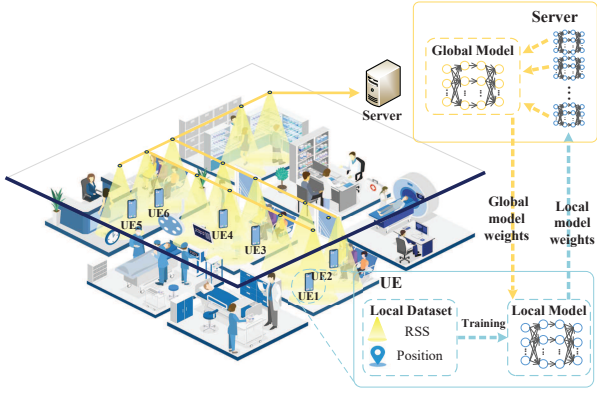}
	\caption{The cooperative VLP framework based on FL.}
	\label{fig:cooperativeSystem}
\end{figure}

To deal with the nonstationary environments, in this paper, a cooperative VLP framework is designed, as illustrated in Figure \ref{fig:cooperativeSystem}. A typical indoor scenario in the hospital is considered as an example, where $N_\mathrm{t}$ LEDs are deployed as access points of both VLC and VLP, and $N_\mathrm{r}$ UEs equipped with abundant computing resources, e.g., mobile phones or intelligent medical devices, are carried by users and move in the building. Using the uplink wireless transmission, such as WiFi and Bluetooth, the UEs can transmit the data of local model weights to the server. Moreover, each UE can store a local dataset and a local model for position estimation, which is feasible for off-the-shelf mobile devices \cite{Ignatov2018ECCV}. In the proposed cooperative VLP framework, UEs participate in the collection of training data and construct their local datasets $\mathbf{\Omega}_j$. Exploiting the local datasets, the UEs cooperatively train a global model applicable for the whole indoor environment using the FL-based algorithm, which is elaborated in the following subsection.

\subsection{FL-Based VLP Algorithm}

\begin{figure}[!t]
	\centering
	\includegraphics[width=0.45\textwidth]{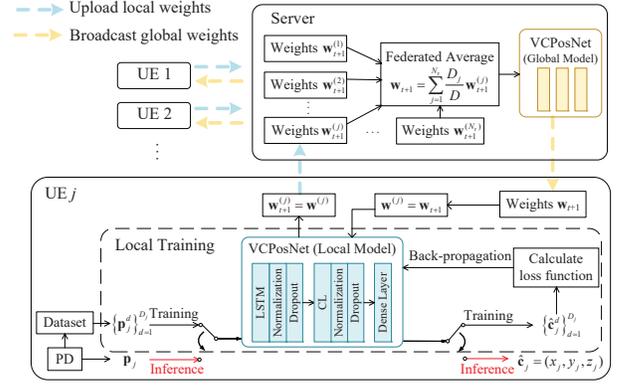}
	\caption{The block diagram of the proposed FL-based VLP algorithm.}
	\label{fig:proposedMethod}
\end{figure}

In this section, we introduce the proposed FL-based VLP algorithm, starting from the multi-user data collection, and then describing the structure of the VCPosNet and the FL-based training and inference procedures. The block diagram of the proposed FL-based VLP algorithm is illustrated in Figure \ref{fig:proposedMethod}.

\subsubsection{Multi-User Sensing and Position Data Collection}

To ensure the samples in local datasets can represent the real-time indoor environment, UEs collect the training samples and update the local datasets over time. Specifically, in a sensing interval, the $j$-th UE measures the received signal power vector $\mathbf{p}_j^d = [P_{1,j}^d, P_{2,j}^d, \cdots, P_{N_\mathrm{t},j}^d]$ transmitted from the $N_\mathrm{t}$ LEDs, and labels them with the corresponding position coordinate $\mathbf{c}_j^d=[x_j^d,y_j^d,z_j^d]$. As users change their locations, the received signal power vector $\mathbf{p}_j^d$ corresponding to different locations are collected by their UEs. Thus, each UE can construct a local dataset $\mathbf{\Omega}_j=\{\mathbf{p}_j^d,\mathbf{c}_j^d \}_{d=1}^{D_j}$ consisting of $D_j$ samples. Unlike the most learning-based VLP methods using a one-shot site survey, in the proposed scheme, data collection is implemented over time. Then, the local dataset is accordingly updated with the new samples, to make it suitable for the real-time environment.

\subsubsection{VCPosNet Structure for Federated Learning}

To better extract the location-related features contained in the received power vector $\mathbf{p}_{j}$ and improve the positioning accuracy, we devise a neural network architecture called VCPosNet in this paper.

In the proposed VCPosNet, firstly, a convolution layer (CL) with $N_\mathrm{f}$ kernel filters is employed, of which the output is applied with the rectified linear units (ReLU) activation function. Then, to prevent possible overfitting because of accumulated parameters, some neurons are dropped out with a probability in the training stage. After this, the resultant data is normalized to ensure that the output data has zero mean and unit standard deviation, to accelerate network training and prevent gradient vanishing. Subsequently, the output data is fed into the long short-term memory (LSTM) layer consisting of $N_\mathrm{u}$ LSTM units, whose output is fed into a Tanh activation function, and also followed by a dropout and a normalization layer. Finally, a fully-connected layer is connected with the final output of the neural network to generate the estimated position coordinate $\hat{\mathbf{c}}_{j}$.

\subsubsection{FL-Based Continuous Model Training and Inference}

To obtain a neural network model adaptive to the nonstationary environments, in this paper, FL is applied to train the VCPosNet in a cooperative manner. Specifically, the training process is composed of the local training processes at the UEs and the federated aggregation of the global model at the server. The detailed procedures are performed as follows and illustrated in \textbf{Algorithm \ref{alg:flTrain}}.

\begin{algorithm}[t!]
	\caption{FL-Based VLP Algorithm (Training Stage)}
	\label{alg:flTrain}
	\begin{algorithmic}[1]
		\REQUIRE ~~\\
		1) Maximum number of local epochs $E$\\
		2) Minibatch size $B$\\
		3) Local datasets $\boldsymbol{\Omega}_1, \boldsymbol{\Omega}_2, \cdots, \boldsymbol{\Omega}_{N_\mathrm{r}}$ \\
		\STATE Initialize the global weights $\mathbf{w}_{0}$
		\FOR{each communication round $t=1,2,\cdots$}
		\STATE \textbf{At UEs:} (parallelly performed by each UE $j$, $\forall j = 1, \cdots N_\mathrm{r}$)
		\STATE Initialize the local weights $\mathbf{w}^{(j)} \leftarrow \mathbf{w}_{t}$
		\FOR{each local epoch $e=0,\cdots,E-1$}
		\STATE Sample a mini-batch $\boldsymbol{\Omega}_{j}^{e}$ from the local dataset $\boldsymbol{\Omega}_j$
		\STATE Compute the loss function $\mathcal{L}(\mathbf{w}^{(j)}, \boldsymbol{\Omega}_{j}^{e})$ by (\ref{eq:loss}) and update the local weights $\mathbf{w}^{(j)}$ using back-propagation
		\ENDFOR
		\STATE Submit the local weights $\mathbf{w}_{t+1}^{(j)} = \mathbf{w}^{(j)}$ and dataset size $D_j$ to server via the uplink transmission
		\STATE ~~\\
		\STATE \textbf{At the Server:}
		\STATE Aggregate the weights $\mathbf{w}_{t+1}^{(1)}, \cdots ,\mathbf{w}_{t+1}^{(N_\mathrm{r})}$ by (\ref{eq:aggregation}) and broadcast $\mathbf{w}_{t+1}$ to all the $N_{\rm{r}}$ UEs via VLC downlink transmission
		\ENDFOR
	\end{algorithmic}
\end{algorithm}

At each UE, the local dataset $\mathbf{\Omega}_j$ is used to train the weights of VCPosNet to obtain a local model containing the environment information of local areas. Mean square error (MSE) is utilized as the loss function, which is given by
\begin{equation} \label{eq:loss}
	\mathcal{L}\left(\boldsymbol{\Omega}_{j}, \mathbf{w}\right)=\frac{1}{D_{j}} \sum_{d=1}^{D_{j}}\left\|\hat{\mathbf{c}}_{j}^{d}-\mathbf{c}_{j}^{d}\right\|_{2}^{2},
\end{equation}
where $\hat{\mathbf{c}}_{j}^{d}=\mathcal{M}(\mathbf{p}_{j}^{d}, \mathbf{w})$ represents the output of the VCPosNet with the input of $\mathbf{p}_{j}^{d}$ and the model weights of $\mathbf{w}$. In each local epoch, the local weights $\mathbf{w}^{(j)}$ are updated using stochastic gradient descent and back-propagation to minimize the loss function using mini-batches $\boldsymbol{\Omega}_{j}^{e}$ of size $B$ sampled from $\boldsymbol{\Omega}_{j}$. After $E$ local epochs, the $j$-th UE uploads the learned local weights $\mathbf{w}_{t+1}^{(j)}$ and the dataset size $D_j$ to the server.

At the server, by applying a weighted average proportional, the local weights $\mathbf{w}_{t+1}^{(1)}, \mathbf{w}_{t+1}^{(2)}, \cdots ,\mathbf{w}_{t+1}^{(N_\mathrm{r})}$ received from UEs are aggregated into the global weight $\mathbf{w}_{t+1}$, which is calculated by
\begin{equation} \label{eq:aggregation}
	\mathbf{w}_{t+1}=\sum_{j=1}^{N_{\mathrm{r}}} \frac{D_{j}}{D} \mathbf{w}_{t+1}^{(j)},
\end{equation}
where $D=D_1+D_2+\cdots+D_{N_\mathrm{r}}$ is the total number of training samples owned by all the UEs. Then, the global weights $\mathbf{w}_{t+1}$ are broadcasted to all the UEs via the VLC downlink transmission.

Afterwards, the $j$-th UE updates its local weight $\mathbf{w}^{(j)}$ with the received global weights $\mathbf{w}_{t+1}$, and then continues the local training processes in the next communication round. With the increase of communication rounds, a temporally generalized position estimation model covering the whole LED-illuminated indoor environment is well trained with the cooperation of multiple users. Moreover, once the indoor environment changes, the global model weights will converge to the optimal values again using updated local datasets in a few communication rounds. Thus, the global model will adapt to the current channel condition rapidly, making it robust to the environmental changes.

Since network training is continuously conducted, we consider that the inference process may happen in any communication round. Once the $j$-th UE requests for the instantaneous position coordinate, the $j$-th UE can measure the received power vector $\mathbf{p}_j$ with PD, then input it into the local model with the latest global weights $\mathbf{w}_t$ to estimate the real-time location coordinate $\hat{\mathbf{c}}_j$.

With the aid of the FL framework and the cooperation of multiple users, an adaptive neural network with robust generalization and adaptation ability can be obtained, which is verified by the simulation results in Section \ref{sec:simulationResults}. In addition, as the training datasets containing massive private data are kept locally, the privacy of users is effectively protected through physical isolation.

\section{Simulation Result and Discussion} \label{sec:simulationResults}

\begin{figure*}[htbp]
	\centering
	\begin{minipage}{0.32\linewidth}
		\centering
		\includegraphics[width=1\textwidth]{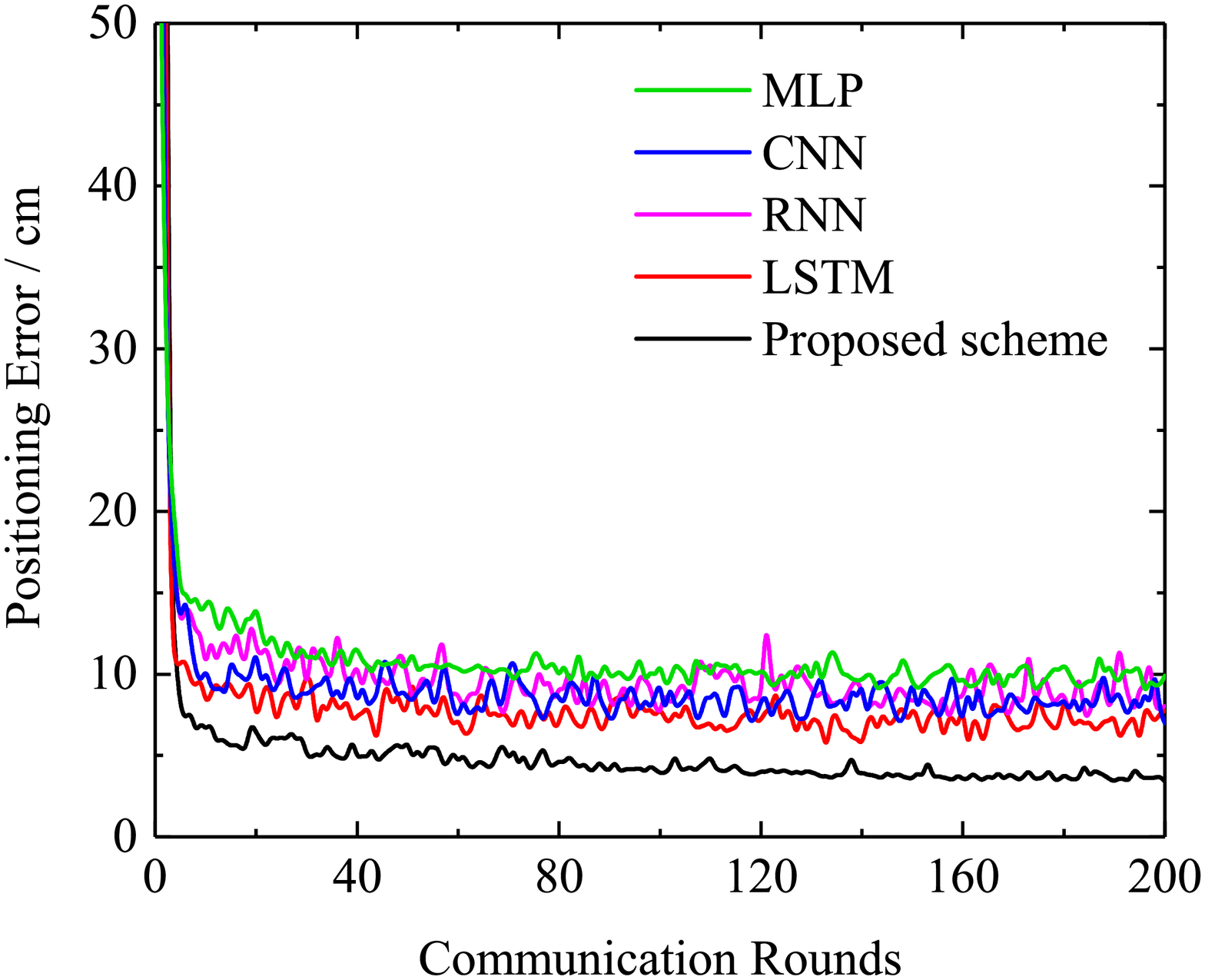}
		\caption{Positioning error with respect to communication rounds.}
		\label{fig:errorCommunRounds}
	\end{minipage} \ \
	\begin{minipage}{0.32\linewidth}
		\centering
		\includegraphics[width=1\textwidth]{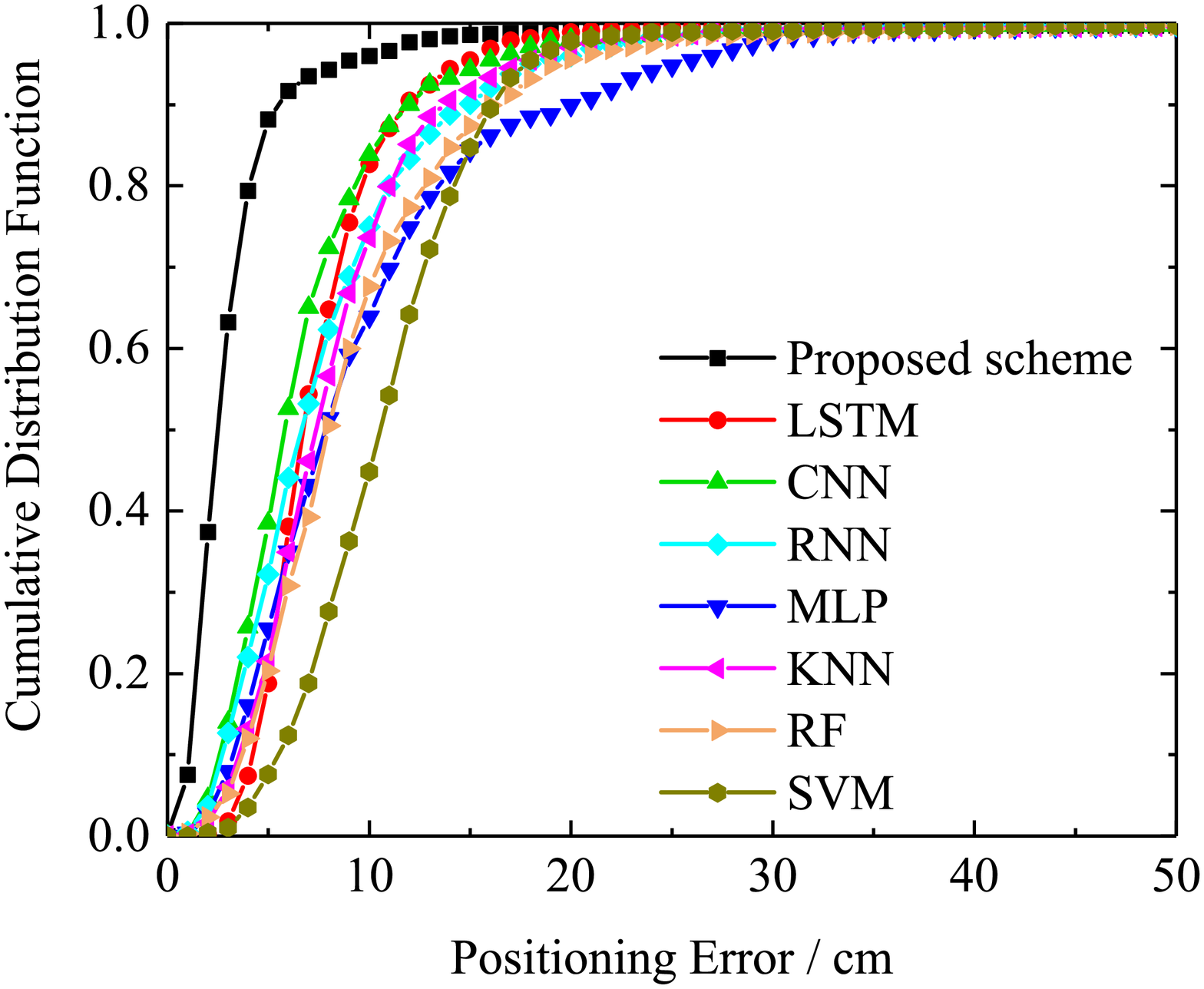}
		\caption{CDF with respect to the positioning error.}
		\label{fig:CDF}
	\end{minipage} \ \
	\begin{minipage}{0.32\linewidth}
		\centering
		\includegraphics[width=1\textwidth]{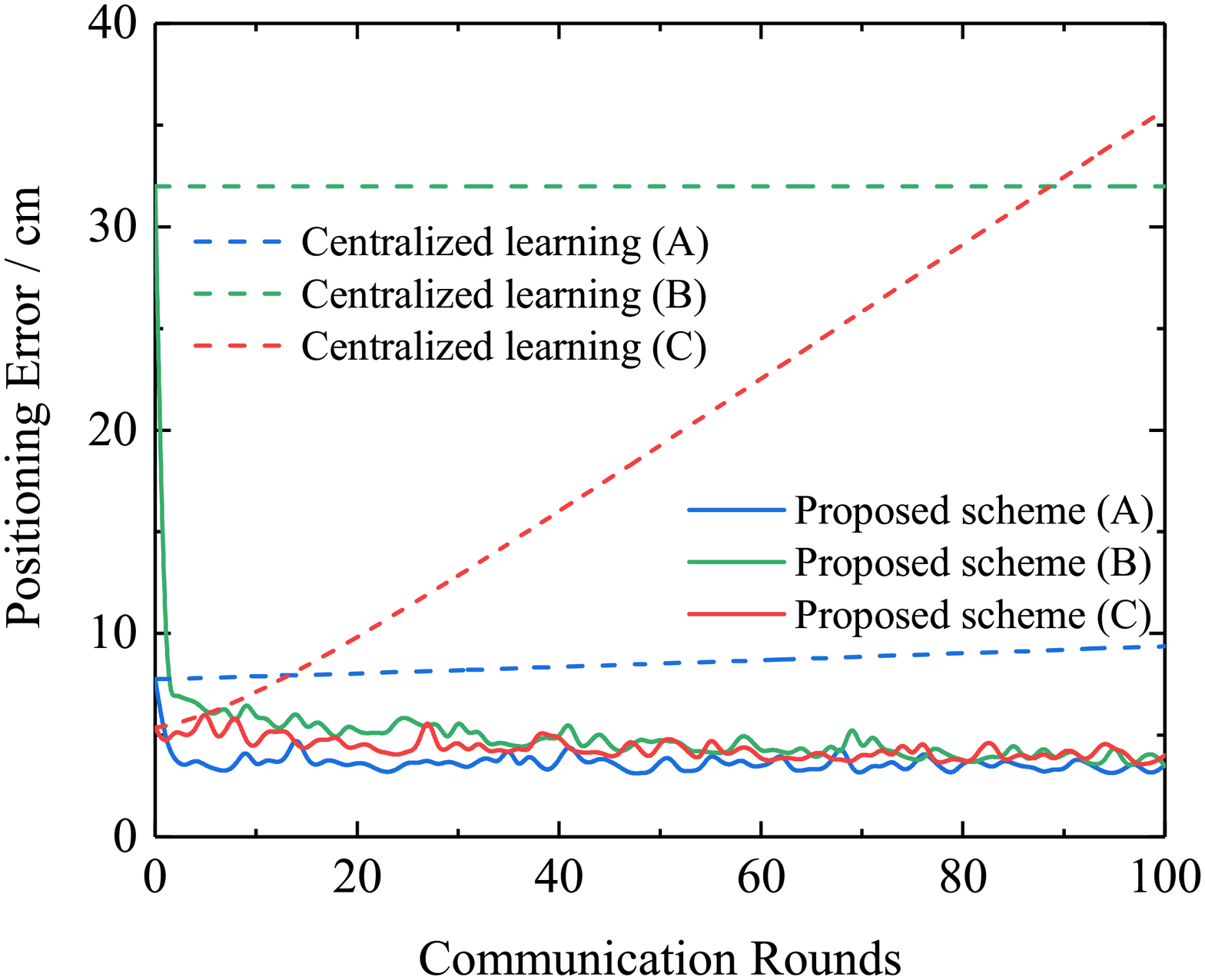}
		\caption{Positioning error in nonstationary environments.}
		\label{fig:nonstationary}
	\end{minipage}
\end{figure*}

In this section, extensive simulations have been conducted to evaluate the performance of the proposed scheme. We consider an indoor scenario in a room with the size of $L \times W \times H=5 \times 5 \times 3 \mathrm{m}^2$, where $N_\mathrm{t}=16$ LEDs are vertically mounted at the inner vertexes of $5 \times 5$ latticed ceiling. The transmitted optical power of the LED is set as $P_i^\mathrm{t}=1\mathrm{W}$, and the half-power angle is $\varphi_{1 / 2}=60^\circ$. The Lambertian order is $m=1$. In addition, the average reflectance of the wall is set as $\bar{\rho}=0.7$. As to the PD, the FOV angle is set as $\theta_\mathrm{FOV}=90^\circ$. The effective area is set as $A_\mathrm{PD}=1 \mathrm{cm}^2$, and the responsivity is set as $R_\mathrm{p}=0.6\mathrm{A/W}$. Moreover, the optical filter gain is set as $T_\mathrm{s}=1$, and the optical concentrator refractive index is set as $n=1$. The background current is set as $I_\mathrm{bg}=740 \mu \mathrm{A}$, and the noise bandwidth is set as $B=5\mathrm{MHz}$. Other parameters are set following literature \cite{Komine2004VLCTCE}. We assume that $N_\mathrm{r}=10$ UEs participate in cooperative model training. Each UE owns a local dataset with the size of $D_\mathrm{r} = 900$, and updates the local dataset with newly collected data every 10 communication rounds. In the training process, the number of local epochs is set as $E=5$, and the minibatch size is set as $B = 128$.

The positioning error of the proposed FL-based VLP scheme with respect to communication rounds is shown in Figure \ref{fig:errorCommunRounds}, compared with benchmark schemes of MLP, CNN, recurrent neural network (RNN), and LSTM trained using the FL framework. It is shown by the results that, the positioning error of the proposed scheme decreases rapidly over time with the cooperation of multiple UEs, and reaches centimeter-level accuracy after only two communication rounds, which verifies the superior convergence capability of the proposed FL-based VLP scheme. After 200 communication rounds, the positioning error of the proposed scheme reaches 3.38 cm, which outperforms the benchmark schemes by 6.52 cm, 3.61 cm, 4.67 cm, and 4.37 cm, respectively.

Subsequently, the cumulative distribution function (CDF) with respect to the positioning error of the proposed scheme is reported in Figure \ref{fig:CDF}, and compared with benchmark methods, including the deep-learning-based methods of MLP, CNN, RNN, and LSTM, and machine-learning-based methods of KNN and support vector machine (SVM). The deep-learning-based methods are trained with the FL framework. The machine-learning-based methods are usually implemented relying on the whole dataset and thus are trained in a centralized manner. It is noted that the CDF of the proposed scheme grows rapidly, and then reaches the value of 0.96 at the positioning error of 10 cm, which means that the positioning accuracy can be controlled within the centimeter-level with a probability of 96\%. Moreover, it can be observed that, the proposed scheme can achieve a higher CDF at the different positioning errors than benchmarks, which verifies the proposed VCPosNet can extract the position-related knowledge effectively and further improve the positioning accuracy.

To further investigate the generalization capability of the proposed method in nonstationary environments, we consider three dynamically changing scenarios, i.e., some parameters of the indoor environment change over time. The details of the three scenarios are as follows. (A) Time-varying ambient light intensity. The background current of the ambient light is initially set as $I_\mathrm{bg} = 0$ before the first communication round, and is increased by $50 \mu \mathrm{A}$ after every communication round. (B) Hardware failure or obstruction blocking. Some LED signals are unavailable to the UEs, i.e., the received signal power from an LED suddenly changes to zero before the first communication round. (C) Device aging. The LED emitting power degrades gradually over time, which is modeled as $P_i^\mathrm{t}=P^{\mathrm{init}} \exp (-\beta t)$, where the initial power $P^{\mathrm{init}}$ is set as 1 W, and the decay coefficient $\beta$ is set as $-\ln(0.8)/100$.

The positioning errors of the proposed FL-based VLP scheme in nonstationary environments are illustrated in Figure \ref{fig:nonstationary}, compared with the centralized-learning-based scheme based on the one-shot site survey. For a fair comparison, the initial model weights of both the proposed scheme and the centralized-learning-based scheme are set as the pre-trained model weights before the first communication round. It is shown by the results that, the positioning accuracy of the centralized-learning-based scheme degrades rapidly with the aggravation of environmental changes. However, in the proposed FL-based VLP framework, the global model weights can adapt to the time-varying channel conditions with the cooperation of multiple users, and the positioning error can be kept at a satisfactory level, which verifies the generalization capability in nonstationary environments.

\section{Conclusion} \label{sec:conclusion}
In this paper, a privacy-preserving cooperative VLP scheme based on FL is proposed, where multiple users cooperatively train a global model adaptive to nonstationary environments. Additionally, The CVPosNet network structure is presented to extract the position-related features and improve the positioning performance. Simulation results show that the proposed scheme outperforms the benchmark schemes in positioning accuracy and generalization performance, especially in the nonstationary environment.

\bibliographystyle{ACM-Reference-Format}
\bibliography{ref}

\end{document}